\documentclass[aps,prd,eqsecnum,notitlepage,nofootinbib,superscriptaddress]
{revtex4-2}

\newcounter{xitem}
\usepackage{bm}
\usepackage{stmaryrd}
\usepackage{amsmath}
\usepackage{amssymb}
\usepackage{graphicx}
\usepackage{textcomp}
\usepackage{calrsfs}
\usepackage{yfonts}
\newcommand{\be}{\begin{equation}}                                              
\newcommand{\ee}{\end{equation}}
\newcommand{\ben}{\begin{equation*}}                                            
\newcommand{\een}{\end{equation*}}
\newcommand{\bea}{\begin{eqnarray}}                                             
\newcommand{\eea}{\end{eqnarray}}

\global\arraycolsep=2pt

\usepackage[english]{babel}
\usepackage[utf8]{inputenc}
\usepackage[T1]{fontenc}
\usepackage{amsmath}
\usepackage{bm}

\DeclareMathOperator{\tr}{tr}

\begin{document}

\title{Perspectives on Quantum Friction, Self-Propulsion, and Self-Torque}

\author{Kimball A. Milton}
\email{kmilton@ou.edu}
\affiliation{Homer L. Dodge Department of Physics and Astronomy,
The University of Oklahoma, Norman, OK 73019}

\author{Nima Pourtolami}
\email{nima.pourtolami@gmail.com}
\affiliation{National Bank of Canada, Montreal, Quebec H3B 4S9, Canada}

\author{Gerard Kennedy}\email{g.kennedy@soton.ac.uk}
\affiliation{School of Mathematical Sciences,
University of Southampton, Southampton, SO17 1BJ, UK}
\date{\today}

\begin{abstract}
  This paper provides an overview of the nonequilibrium fluctuational forces
  and torques acting on a body either in motion or at rest relative to 
another body or to the thermal vacuum blackbody radiation. 
We consider forces and torques beyond the usual static Casimir-Polder and
Casimir forces and torques.
For a moving body, 
  a retarding force emerges, 
   called quantum or Casimir friction,  which in vacuum was first predicted
   by Einstein and Hopf in 1910. Nonreciprocity may allow
a stationary body, out of thermal  equilibrium with its environment, to 
experience a torque.
 Moreover, if a stationary reciprocal body is not  
   in  thermal equilibrium with the blackbody vacuum, a self-propulsive 
   force or torque can appear, resulting in a potentially observable linear 
   or angular terminal velocity, even after thermalization.
\end{abstract}

\maketitle

\section{Introduction}

   For nearly 50 years, it has been appreciated that when two bodies
  move parallel to each other, they experience a quantum
  frictional force that tends to retard the relative motion
  \cite{Teodorovich}.
  But more than a century ago, it was recognized \cite{EH} that a moving 
  body experiences friction even in vacuum!   
  Even more remarkably, in the last decade it has been shown that
 a stationary body out of thermal equilibrium with the background blackbody 
 radiation
 will experience a force and a torque, if it is suitably asymmetric.
 Here we will discuss such phenomena, which are on the verge of
 observability, in a perturbative context, expanding in $\varepsilon-1$,
 where $\varepsilon$ is the permittivity of the object.  A brief perspective
 of this field of research is provided in context.
 
  We use natural units, $\hbar=c=\epsilon_0=\mu_0=k_B=1$, and Heaviside-Lorentz
units for polarizability, which is related to the Gaussian polarizability by
$\alpha^{\rm HL}=4\pi \alpha^{\rm G}$.

\section{Friction of particle with a dielectric or metallic surface}
Casimir friction refers to the dissipative retarding
force experienced by a body (an atom, a nanoparticle, 
or a dielectric plate) moving parallel to another body (typically
a metallic or dielectric plate).  The subject has a long history, as noted,
but the literature can be confusing because
there is considerable controversy surrounding
both the velocity and distance dependence, as well as the effect of
temperature. There is no real  experimental evidence for Casimir
friction, although 
hundreds of theoretical papers on the subject
 have been published.  For a review of
the literature a decade ago see Ref.~\cite{rev}.
For a more recent detailed perspective on atom-surface friction, 
with many references, see Ref.~\cite{persp22}

\begin{figure}
\begin{minipage}{.45\textwidth}
\centering
\includegraphics[width=\textwidth, trim = -3cm 20cm 10cm 1cm, clip]{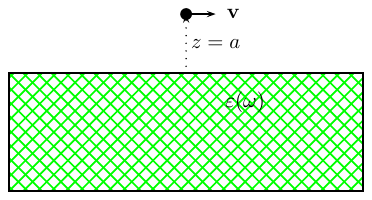}
\caption{
A particle moving with velocity $v$ in vacuum 
a distance $a$ above a dielectric plate.
Polarizability is defined by $\mathbf{d}=\bm{\alpha}\cdot \mathbf{E}$.}
\label{fricwdiel}
\end{minipage}\hfill
\begin{minipage}{.45\textwidth}
\includegraphics[width=.8\textwidth]{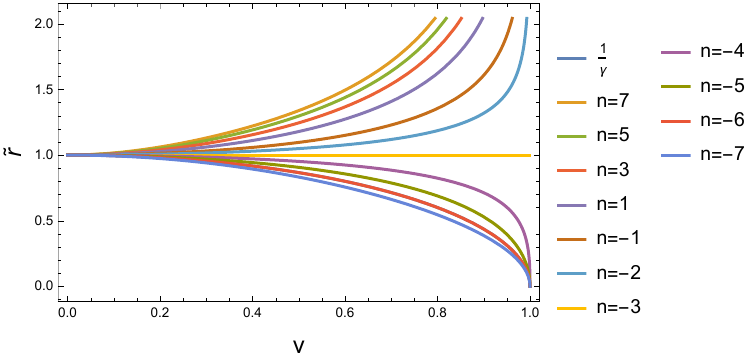}
\caption{The ratio
$\tilde r=\frac{\tilde T}{T}$, versus velocity $v$,
where $T$ is the temperature of the blackbody
radiation background, and $\tilde T$ is the temperature of the particle in 
NESS, where the particle neither gains nor loses energy.  (The dynamic 
version of thermal equilibrium.)  
If the temperature ratio could be measured, that would be
a signal of quantum friction. Note that $n=3$ is the pure radiation reaction
model (\ref{rr}), with $\alpha_0$ constant. The temperature ratio is unity 
for $n=-3$, and is exactly $1/\gamma=\sqrt{1-v^2}$ for $n=-6$.}
\label{mononess}\end{minipage}
\end{figure}

Here, let us consider  a particle,
described by a static
electric  polarizability $\alpha_0$, moving with velocity $v$ 
parallel to a
metallic plate  with conductivity $\sigma$, a distance $a$ above the plate,
as illustrated in Fig.~\ref{fricwdiel}.
The origin of the theoretical controversy is that different physical
mechanisms are considered as responsible for  the dissipative effect
of the polarizable particle.  
For an atom, the most significant mechanism arises from the resistance 
in the metal plate, because the image in the metal
of the charge distribution of the moving polarizable particle
lags behind that of the particle itself.
This gives rise, on use of the
fluctuation-dissipation theorem [see Eq.~(\ref{fdteqns}) below], to 
a frictional force on the particle that is \cite{Intravaia,Hoye},
nonrelativistically, at zero temperature,
\begin{equation}
    F=-\frac{135 \alpha_0^2 v^3}{2\pi^3\sigma^2(2a)^{10}}.\label{fimage}
\end{equation} 
If, in contrast, one were to assume that radiation reaction 
[see Eq.~(\ref{rr}) below] were
the dominant mechanism by which the polarizability of the particle acquires
a dissipative part, the force would be considerably different \cite{Hoye},
\begin{equation}
F=-\frac{105}{128\pi^3}\frac{\alpha_0^2 v^5}{\sigma a^9}.\label{frr}
\end{equation}
In addition, the nanoparticle itself (not an atom) might well possess 
intrinsic dissipation of its own, say if it were composed of a realistic
metal.  
If this were the only effect, and the conductivity of the nanoparticle
were $\sigma'$, while
the plate had conductivity  $\sigma$, 
 the force would be first order in the
static polarizability of the particle \cite{nomark},
\begin{equation}
F=-\frac{135}{64\pi^2}\frac{\alpha_0}{\sigma\sigma'}\frac{v^3}{a^7}.\label{fo}
\end{equation}
(These results have been recalculated using the general reformulation
of quantum friction in Refs.~\cite{guoqf1,guoqf2}, 
so there are some discrepancies in
the numerical coefficients found in the original literature.)
All these effects, in principle, are present, and it can be difficult
to disentangle their importance, especially in the absence of any experimental
guidance.  We note that the force (\ref{fimage}) dominates 
the radiation-reaction force (\ref{frr}) at low velocities
and short distances; the two expressions become comparable at a separation of
10 nm if $v$ is of order $10^6$ m/s. If present, the first-order effect
for a metallic nanoparticle (\ref{fo}) overwhelmingly dominates.
The reader will note that these results are all given for zero temperature.
This is because then the integrations are restricted to small values of
frequency, where only the TM reflection coefficient need be considered, and
the integrations become trivial.  The general case of nonzero temperature
is considerably more subtle, but less relevant in practice.

Let us  give a brief survey of some of the novel features of Casimir
friction with surfaces explored during the last few years.  Friction may 
occur on a nanosphere moving parallel to a dielectric surface, if it is 
moving faster than the speed of light in that medium
(\v{C}erenkov friction) \cite{Silveirinha,Pieplow}.  
There is extensive literature
concerning the friction between parallel moving plates, for which we cite
Refs.~\cite{Dedkov20,BSS}, for example.  A rotating particle near an
isotropic surface may experience a frictional torque \cite{Zhao}.  
Exotic metamaterials may induce forces on rotating nanoparticles \cite{Wang}.
Nonequilibrium thermodynamics comes into play in quantum friction
\cite{reiche,IB}, as we shall see.
A covariant approach to radiative friction of a particle above a surface and
in vacuum is given in Ref.~\cite{PH}.  A very nice summary of the theory of
quantum friction near surfaces is given in the thesis \cite{marty}.
 
 However, the most important 
issue in the field
is the lack of experimental input.  The quantum forces tend to be exceedingly
small, so novel methods of finding signatures of friction must be sought.
One exciting possibility is looking for accumulated geometric phase
\cite{Farias}; another scheme, reflecting the nonequilibrium nature of
quantum friction, is explored in Sec.~\ref{sec2}.  There is even a report
\cite{Xu} of experimental observation of quantum friction, although in an
analog mechanical system.

\section{Quantum vacuum friction---Einstein-Hopf formula}
\label{sec2}
 If a particle has dissipation (which it must, if only through interaction
 with the electromagnetic field), it will experience quantum vacuum friction,
 which, nonrelativistically, is given by
the Einstein-Hopf formula \cite{EH},
\begin{equation}
F=-\frac{v\beta}{12\pi^2}\int_0^\infty d\omega\,\omega^5
\Im\alpha(\omega)\frac1{\sinh^2\beta\omega/2},\label{eh}
\end{equation}
which is linear in the velocity,
where $\beta=1/T$ refers to the temperature.
Here, for example, if the dissipation occurs entirely from radiation reaction,
as appropriate for an atom,
\begin{equation}
    \Im \alpha(\omega)=\frac{\omega^3}{6\pi}\alpha_0^2(\omega),\label{rr}
\end{equation}
where $\alpha_0(\omega)$ is the real, nondissipative, intrinsic polarizability
of the atom.

For low temperature ($<10^6$ K), low velocity ($v\ll c$), and if
dispersion is neglected, this frictional
force reduces to
\begin{equation}
    F(v)=-\frac{32 \pi^5\alpha_0^2}{135 \beta^8}v=m\frac{dv}{dt},
    \end{equation}
    which means the time required for a atom to slow from an initial
    velocity $v_i$ to a final velocity $v_f$ is
    \begin{equation}
\Delta t=-t_0\ln \frac{v_f}{v_i},\quad t_0=
\frac{135m\beta^8}{32\pi^5\alpha_0^2}=1.7 \times 10^{25}\,\mbox{s}
    \end{equation}
for gold at $T=300$ K.   For a 10\% decrease in velocity, $\Delta t$   
 would drop to the lifetime of a US physics
graduate student, 5.9 years, if the temperature could be raised to 30,000~K!

 For higher velocities, we need to distinguish between the temperature, $T$,
of the blackbody background,  and the temperature, $T'$, of the body.  There
is a special ratio of $T'/T$, which is the generalization of thermal
equilibrium, in which the body neither gains nor loses energy in its
rest frame.  This is called the ``nonequilibrium steady state'' (NESS)
\cite{reiche}. This
ratio is illustrated in Fig.~\ref{mononess} for a monomial dissipation law,
that is, where $\Im\alpha(\omega)\propto \omega^n$.  Measuring this
temperature ratio might be a way of finding a signal of quantum friction,
but would require velocities comparable to the speed of light.  For more
detail on relativistic quantum vacuum friction 
at arbitrary temperatures see \cite{guoqf1,guoqf2}.
This, of course, was discussed earlier, for example in 
Refs.~\cite{volokitin,Dedkov,VP}, which also discuss the NESS temperature.

\section{Quantum vacuum torque: nonreciprocal media}
\label{sec:nrtorque}
Classically, the torque on a stationary dielectric
body with polarization vector $\mathbf{P}$ is
given by \cite{CE}
\begin{equation}
\bm{\tau}=\int (d\mathbf{r})\frac{d\omega}{2\pi}\frac{d\nu}{2\pi}
e^{-i(\omega+\nu)t}\left[\mathbf{P(r};\omega)\times \mathbf{E(r};\nu)+
P_i(\mathbf{r};\omega)(\mathbf{r}\times\bm{\nabla})E_i(\mathbf{r};\nu)
\right]\!.\label{torqueform}
\end{equation}
The first term here is called the {\it internal\/} torque and the second the
{\it external\/} torque, because the latter is reflective of the force on the body. Now we use the first-order expansions 
\begin{equation}
\mathbf{E}^{(1)}(\mathbf{r};\omega)=\int (d\mathbf{r'})\bm{\Gamma}(\mathbf{r-r'};\omega)\cdot
\mathbf{P}(\mathbf{r}';\omega),\quad
\mathbf{P}^{(1)}(\mathbf{r};\omega)=\bm{\chi}(\mathbf{r};\omega)
\cdot\mathbf{E}(\mathbf{r};\omega).\label{order1exp}
\end{equation}
We evaluate the EE and PP contributions to the torque by use of the 
fluctuation-dissipation theorem (FDT):
\begin{subequations}
\label{fdteqns}
\begin{eqnarray}
\langle \mathbf{P}_i(\mathbf{r};\omega)\mathbf{P}_j(\mathbf{r}';\nu)\rangle&=&
2\pi\delta(\omega+\nu)\delta(\mathbf{r-r'})\bm{\chi}^A_{ij}(\mathbf{r};\omega)\coth
\beta'\omega/2,
\\
\langle\mathbf{E}_i(\mathbf{r};\omega)\mathbf{E}_j(\mathbf{r}';\nu)\rangle&=&
2\pi\delta(\omega+\nu)\Im\bm{\Gamma}_{ij}(\mathbf{r-r'};\omega)\coth
\beta\omega/2,
\end{eqnarray}
\end{subequations}
where symmetrization is understood for the field products,
$T=1/\beta$ is the blackbody
vacuum temperature, and $T'=1/\beta'$ is the body temperature.
Here, $\bm{\Gamma}$ is taken to be the usual vacuum retarded Green's dyadic, which at
coincident points is (rotationally averaged)
\begin{equation}
    \bm{\Gamma}(\mathbf{r-r'};\omega)\to\bm{1}\left(\frac{\omega^2}{6\pi R}+i\frac{\omega^3}{6\pi}+O(R)\right),\quad R=|\mathbf{r-r'}|\to0,
\end{equation}
while $\bm{\chi}^A$ is the anti-Hermitian part of the electric 
susceptibility:
\begin{equation}
    \bm{\chi}^A=\frac1{2i}(\bm{\chi}-\bm{\chi}^\dagger), \qquad
    \Im\bm{\chi}^A_{ij}=-\frac12\Re[\chi_{ij}(\omega)-\chi_{ji}(\omega)],
\end{equation} 
where we recognize that
only the antisymmetric part of $\bm{\chi}^A$, which is even 
in $\omega$, contributes to the torque.  The real part of $\bm{\chi}^A$ is 
symmetric in indices and odd in $\omega$. $(\Im\bm{\chi}^A=0$ for a 
reciprocal body).

Using this, we readily calculate \cite{torque1,GK}
the torque on a nonreciprocal body in vacuum, 
due to 
PP and EE fluctuations, which arises only from the internal torque,
\begin{equation}
    \tau_i=-\int_{-\infty}^\infty \frac{d\omega}{2\pi} \frac{\omega^3}{6\pi}\left[
    \coth\frac{\beta'\omega}2-\coth\frac{\beta\omega}2\right]\epsilon_{ijk}\Re\alpha_{jk}
    (\omega),
   \qquad \alpha_{jk}(\omega)=\int(d\mathbf{r})\chi_{jk}(r;\omega).
\end{equation}
The system must be out of thermal equilibrium for a torque to occur.
This result exactly agrees with that of Refs.~\cite{fan,strekha}.  
Closely related is the observation that an external magnetic field may
result in such a torque on a stationary spherical particle \cite{Pan}, which 
makes explicit the nonreciprocity arising from the applied field 
\cite{torque1,GK}.
However, there is no quantum
vacuum force in this static situation.
A nonreciprocal medium typically requires an external magnetic
field. If the nonreciprocity is generated by $B=1$\,T, the corresponding torque for a gold  ball of radius 100\,nm is $\sim 10^{-24}$\,N\,m, for $T'/T=2$.

A topological insulating film in a magnetic field can also experience a
nonequilibrium torque \cite{maghrebi}.
The angular
 momentum flux radiated by a magneto-optical nanoparticle, an example
of a nonreciprocal body, was discussed in Ref.~\cite{Ott}.
Forces and torques can also be induced on a particle or body in the
near field of a surface made nonreciprocal by the presence of a magnetic field \cite{khandekar,gelbwasser}.

\section{Self-Propulsive Force}
The Lorentz force on a dielectric body is \cite{CE}
\begin{equation}
\mathbf{F}=\int(d\mathbf{r})\int \frac{d\omega}{2\pi}\frac{d\nu}{2\pi}
e^{-i(\omega+\nu)t}
\bigg\{-\left(1+\frac\omega\nu\right)
\left[\bm{\nabla}\cdot \mathbf{P}(\mathbf{r};\omega)\right]
\mathbf{E}(\mathbf{r};\nu)
-\frac\omega\nu
\mathbf{P}(\mathbf{r};\omega)\cdot(\bm{\nabla})\cdot
 \mathbf{E}(\mathbf{r};\nu)\bigg\}.\label{Lorentzforce}
\end{equation}
Now we expand the fields out to second order (4th order in generalized
susceptibilities, $\bm{\chi}$ and $\bm{\Gamma}$, upon use of FDT), by 
iterating the expansion in Eq.~(\ref{order1exp}).
Then using the symmetries of the integrand we find the general
expression for the force on a body composed of isotropic material.
It is evident that, in this order, there is no force on a homogeneous
body.
\begin{figure}
\centering
\includegraphics[width=.2\textwidth,trim = 1cm 19cm 14cm 1cm, clip]{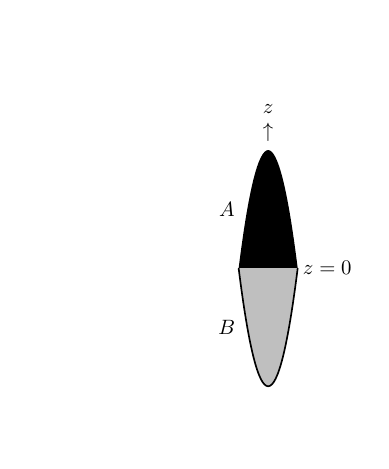} \hfill
\includegraphics[width=.3\textwidth, trim = 1cm 19cm 11cm 3cm, clip]{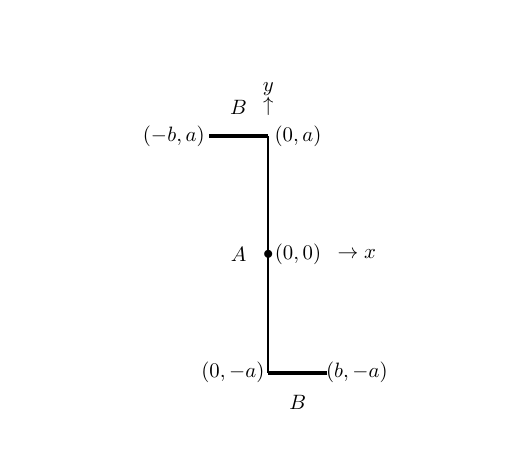}\hfill
\includegraphics[width=.3\textwidth, trim = 1cm 19cm 12cm 2cm, clip]
    {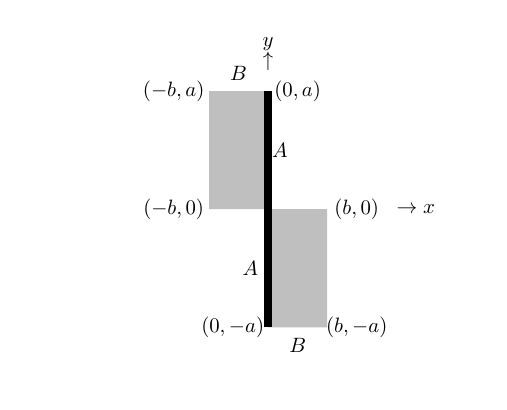}
\caption{ (a) Generic object with two parts. Axial symmetry is shown
for simplicity, so the force is in the $z$ direction.
(b) An inhomogeneous wire of small cross section bent in the shape of 
a dual Allen wrench. The end pieces (``tags'') $B$ are taken to be 
dispersionless dielectric, while the central wire $A$ is a Drude-type metal. 
The Cartesian coordinates of the
various junctions are shown, as is the center of mass.
(c) Dual flag, consisting of a metal central wire, with dielectric
    flags attached antisymmetrically.}
\label{2partbody}
\end{figure}
So, we consider the simplest example of an inhomogeneous body, one
composed of two homogeneous parts, $A$ and $B$, 
 as shown in Fig.~\ref{2partbody}a.
Then, the force reduces to \cite{selfprop}
\begin{equation}
F_z=8\int_0^\infty
\frac{d\omega}{2\pi}X_{AB}(\omega)I_{AB}(\omega)
\left[\frac1{e^{\beta\omega}-1}-\frac1{e^{\beta'\omega}-1}\right],
\end{equation}
where the geometric integral is ($\mathbf{R=r-r'}$, $\tilde{R}=\omega R$)
\begin{equation}
I_{AB}(\omega)=
\int_A(d\mathbf{r})\int_B(d\mathbf{r'})\,
\frac12\nabla_z\left[\Im \Gamma_{ji}(\mathbf{r'-r};\omega)
\Im \Gamma_{ij}(\mathbf{r-r'};\omega)\right]=\int_A(d\mathbf{r})
\int_B(d\mathbf{r'})\frac1{(4\pi)^2}
\frac{R_z}{R^8}\phi(\tilde{R}),\label{IAB}
\end{equation} 
in terms of  the function $\phi$,
\begin{equation}
\phi(\tilde{R})=-9-2\tilde{R}^2-\tilde{R}^4+(9-
16\tilde{R}^2+3\tilde{R}^4)\cos2\tilde{R}+\tilde{R}(18-
8\tilde{R}^2+\tilde{R}^4)\sin2\tilde{R};
\quad \phi(\tilde{R})\sim -\frac49 \tilde{R}^8+\frac{28}
{225}\tilde{R}^{10}+\cdots,
\quad \tilde{R}\ll1.\label{phi}
\end{equation}
The susceptibility product is
\begin{equation}
X_{AB}(\omega)
=\Im\chi_A^{\vphantom{q}}(\omega)
\Re\chi_B^{\vphantom{q}}(\omega)
-\Re\chi^{\vphantom{q}}_A(\omega)
\Im\chi_B^{\vphantom{q}}(\omega).
\end{equation}

In order to apply our weak-susceptibility approximation to a metal,
described by the Drude model, 
\be
\chi(\omega)=-\frac{\omega_p^2}{\omega(\omega+i\nu)},\quad \omega_p \mbox{
is the plasma frequency, and} \,\,\nu\,\, \mbox{the damping frequency},
\ee
we will assume that the object is no thicker than the 
skin depth, which is approximately given by \cite{CE} 
\begin{equation}
\delta=(\omega \sigma/2)^{-1/2}=(\omega^2\Im\chi/2)^{-1/2}
=\sqrt{\frac{2(\omega^2+\nu^2)}{\omega\omega_p^2\nu}}\sim 50 \,\mbox{nm},
\end{equation}
putting in parameter values for gold.

\begin{figure}
\centering
\begin{minipage}{.45\textwidth}
\includegraphics[width=7cm]{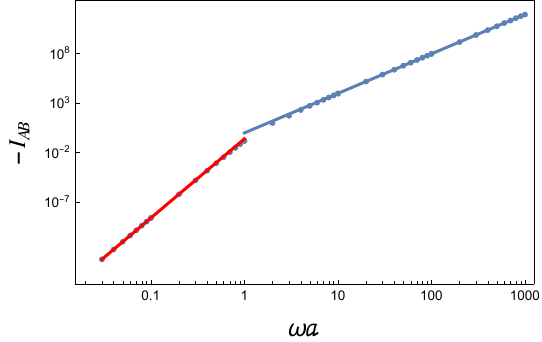}
\caption{Geometric integral for a Janus ball, multiplied by $8\pi a$. The
lines show the large $\sim(\omega a)^4$ and small $\sim( \omega a)^8$
behaviors.}\label{jbi}
\end{minipage}\hfill
\begin{minipage}{.45\textwidth}
\includegraphics[width=7cm]{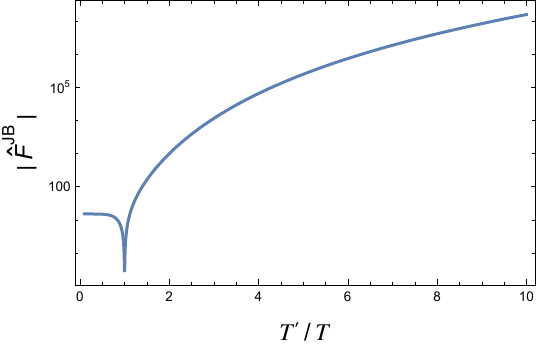}
\caption{Force on Janus ball.  Results are
comparable with those of Ref.~\cite{Reid}, but scaling with $a$ is
different.  The force is negative, that is, toward the metal side, if $T'>T$.}
\label{jbforce}
\end{minipage}
\end{figure}

As an example, consider a Janus ball, of radius $a$,
with one half being a dispersionless
dielectric, and the other half a Drude metal.  In that case, the 
geometric integral $I_{AB}$ is shown in Fig.~\ref{jbi}.
The spontaneous force on a small Janus ball ($A$ being a
dispersionless dielectric and $B$ gold) is
\begin{equation}
F^{\rm JB}=\frac1{27\pi}\chi_A\omega_p^2(\nu a)^7 \hat{F}\approx
4 \times 10^{-25} \chi^{\vphantom{q}}_A\hat{F} \,\mbox{N},
\quad 
\hat{F}=f_7(\beta\nu)-f_7(\beta'\nu),\end{equation}
for a ball of radius 100 nm. Here, the dimensionless force $\hat {F}$
is given in terms of
\begin{equation}
f_n(y)\equiv\int_0^\infty dx\frac{x^n}{x^2+1}
\frac1{e^{y x}-1}.\label{fn}
\end{equation}
$\hat{F}$ is shown as a function of the temperature of the ball in 
Fig.~\ref{jbforce} for $T=300$ K.  The spontaneous force on a Janus
ball was first considered, perturbatively, in Ref.~\cite{muller}, and
nonperturbatively, in Ref.~\cite{Reid}.

A considerably larger force might arise on a planar structure, one
half being absorbing material and the other being a metal.  This was
studied first in Ref.~\cite{Manjavacas} and later in Ref.~\cite{selfprop}.

\section{Thermal relaxation leads to terminal velocity}
Once the body starts to move due to the nonequilibrium quantum force,
it will experience quantum friction, but
this is very small.
More important than quantum friction is thermalization.
Unless some mechanism is provided to maintain the temperature difference, the body will eventually cool or heat to the temperature of the
environment.
From Newton's law, the terminal velocity is, if we assume 
an adiabatic change in the temperature,
\begin{equation}
v^{\vphantom{q}}_T=\frac1m \int_0^\infty dt\, F(T'(t), T).\label{tervelcool}
\end{equation}
Here the cooling rate is given by, for a slowly moving body,
\begin{equation}
    \frac{dQ}{dt}=C_V(T')\frac{dT'}{dt}=P(T',T),\quad
    P(T',T)= \frac1{3\pi^2}\int_0^\infty d\omega\,\omega^4\Im
\tr\bm{\alpha}(\omega)\left[\frac1{e^{\beta\omega}-1}-\frac1{e^{\beta'\omega}-1}\right],\end{equation}
where $C_V(T')$ is the specific heat of the body at temperature $T'$.
In terms of $u=T'/T$, the time taken for a homogeneous
body to cool from temperature
$T_0'$ to $T_1'$ is, for $T_0'>T_1'>T$,
\begin{equation}
t_1=\int_{T_0'}^{T'_1} dT'\frac{C_V(T')}{P(T',T)}, \quad \mbox{or}\quad
   \frac{t_1}{t_c}=\int_{u_0}^{u_1} \frac{du}{p(u,T)}, \quad
p(u,T)=f_3(\beta\nu)-f_3(\beta'\nu),\quad
    t_c=\frac{3\pi^2 n T}{\nu^3\omega_p^2}\sim 10^{-4} \,\mbox{s},
    \end{equation}
    where $n$ is the number density.
Here, we've used the weak susceptibility model of a metal, and
 the assumption that $T\gg\Theta_D$, the Debye temperature.
 (The latter approximation is well-satisfied for $T=300$ K.) 
This means we can write the terminal velocity as
\begin{equation}
    v^{\vphantom{q}}_T=\frac{t_c}m \int_{u_0}^1 du \frac{F(u,T)}{p(u,T)}.
\end{equation}
The terminal velocity turns out to be only about 0.1 nm/s for a Janus ball 
of radius 100 nm, half made of gold and half dielectric, initially twice as 
hot as the background.  Since this seems impossible to  observe directly, in
the next Section we turn to a more accessible possibility.

\section{Spontaneous torque on an inhomogeneous chiral body}

In Sec.~\ref{sec:nrtorque}, 
we calculated the torque in first order, which required the body
be composed of nonreciprocal material, which usually necessitates an external
field be applied.  In second order, a torque can arise
for an ordinary (reciprocal) body, but again only if the
body is {\it inhomogeneous}.  It must further be {\it chiral}, in that any mirror reflection cannot be turned into the original object by translations
or rotations.
For a body with isotropic but inhomogeneous susceptibility there is only
an external torque.
If the body, again, consists of two homogeneous parts, $A$ and $B$, 
 the external torque (\ref{torqueform}) yields \cite{sptorque}
\begin{equation}
\bm\tau=\frac1{2\pi^2}\int_0^\infty \frac{d\omega}{2\pi}X_{AB}(\omega)
\left(\frac1{e^{\beta\omega}-1}-\frac1{e^{\beta'\omega}-1}\right)
\mathbf{J}_{AB}(\omega),\quad \mathbf{J}_{AB}(\omega)=-\int_A(d\mathbf{r})\int_B(d\mathbf{r'})
\frac{\mathbf{r\times r'}}{|\mathbf{r-r'}|^8}\phi(\tilde{R}),
\label{gf}
\end{equation}
where the function $\phi$ is defined in Eq.~(\ref{phi}).

\begin{figure}
\begin{minipage}{.45\textwidth}
\begin{center}
\includegraphics[width=7cm]{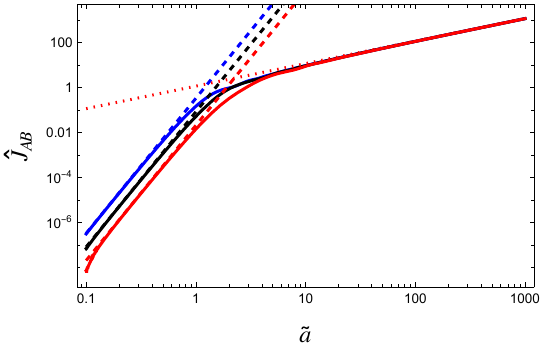}
\end{center}
\caption{Geometric factor for dual Allen wrench, $J_{AB}=2\omega^4 
S_AS_B \hat{J}_{AB}$, for different values of $b=a/2$, $b=a$, $b=2a$ from 
bottom to top. Here $\tilde{a}=a\omega$. }
\label{dawgeo}\end{minipage}\hfill
\begin{minipage}{.45 \textwidth}
\begin{center}
\includegraphics[height=4.5cm]{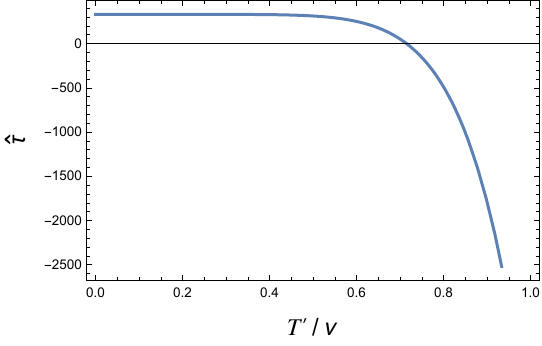}
\caption{Dimensionless torque on a small Allen wrench, for $T=300$ K and
dissipation appropriate for gold.
\\ \hfill\\ \hfill}
\label{saw}    
\end{center}
\end{minipage}
\end{figure}
We consider a ``dual Allen wrench'' as illustrated in Fig.~\ref{2partbody}b.
The object will experience a quantum vacuum torque, perpendicular to its
plane, but not a net force, because it is reflection invariant in the 
center of mass, $\mathbf{r\to-r}$.
The geometric factor is
\begin{equation}
J_{AB}(\omega)=2 S_AS_B\int_{-a}^a dy \int_0^b dx \, x y \frac{\phi(\tilde{R})}{\tilde{R}^8},\quad \tilde{R}=\omega R,
\quad R=\sqrt{x^2+(a+y)^2}.
\end{equation}
In terms of $\tilde{a}=\omega a$, $J_{AB}$ is shown in Fig.~\ref{dawgeo}, 
apart from a prefactor of $2\omega^4 S_A S_B$, where $S_i$ is the 
cross-sectional area of the $i$th wire. 
We see saturation in $\tilde{b}$, and linear behavior in $\tilde{a}$
for large $\tilde{a}$, which is easily understood. The interactions 
between the parts are local, so increasing $b$ beyond a certain point does 
not increase the torque.  The local forces on $A$ are also saturated as $a$ 
increases, but the lever arm increases linearly. 
The asymptotic values of $\hat{J}_{AB}$ are
\begin{equation}
\hat{J}_{AB}\sim \frac{11}{30}\pi\omega a, \quad \omega a\gg1;
\qquad  \hat{J}_{AB}\sim \frac{56}{675}\omega^6 a^4 b^2,
\quad\omega a, \omega b\ll1.
\end{equation}
Around room temperature, the transition from large to small occurs at 
$a$, $b\sim 10$ $\mu$m.

    The situation is most favorable for a small object.  
    The corresponding torque is 
    \begin{equation}
\tau=\frac{28}{675\pi^3}\chi_B \nu^9\omega_p^2S_AS_B a^4b^2\hat{\tau}\equiv
\tau_0\hat{\tau},\quad \hat{\tau}=f_9(\beta\nu)-f_9(\beta'\nu),    
\end{equation}
in terms of the function defined in Eq.~(\ref{fn}), 
where $\hat\tau$ is shown in Fig.~\ref{saw}.
The moment of inertia of the object is
\begin{equation}
I=\rho_A S_A\frac23a^3+\rho_BS_B2b\left(a^2+\frac13 b^2\right),
\end{equation}
so the resulting terminal angular velocity due to thermal cooling is
\begin{equation}
    \omega_T=\frac{t_c\tau_0}I \hat{\omega}_T, \quad \hat{\omega}_T=
    \int_{T'_0/T}^1 du\frac{\hat{\tau}(u,T)}{p(u,T)}, 
\end{equation}
where the prefactor $t_c\tau_o/I \sim 2\times 10^{-7}$  s$^{-1}$ for $a,b$ 
of order 1 $\mu$m, with cross-sectional radius about $50$ nm. Because the 
dimensionless torque is large, so is
$\hat{\omega}_T\sim 20,000$, if the initial temperature of the object
is twice that of the room temperature environment, which leads to a large 
terminal angular velocity
    $\omega_T\sim 4\times 10^{-3}\, \mbox{s}^{-1}$.
This should be quite observable.   Further enhancement, by a factor of
10 or so, occurs if the tags in the dual Allen wrench are unfurled into flags,
as shown in Fig.~\ref{2partbody}c. Many more details are supplied in 
Ref.~\cite{sptorque}.

    So, seeking to observe the torque
on a small chiral object  seems a promising experimental direction.

\section{Conclusions and Perspective}
After briefly reviewing quantum friction with a surface, or with 
vacuum blackbody radiation, we turned to consideration of spontaneous
torques and forces in vacuum.
In first order in electric susceptibility, a vacuum torque, but no
force, can arise for a body made of {\it nonreciprocal\/} material,
if the body is out of equilibrium with the blackbody vacuum environment.
In second order, a vacuum force can arise only if the body is {\it 
inhomogeneous}, but no exotic material properties are required.
A vacuum torque can also arise for ordinary {\it chiral\/}
bodies in second order, 
but again only if the body is also {\it inhomogeneous}.  This is in contrast 
to the nonperturbative findings of Ref.~\cite{Reid}, which discrepancy
is resolved by considering third-order effects (work in progress).
 We consider some examples of bodies which should exhibit 
possibly observable vacuum forces and torques, although cooling (or heating)
to equilibrium
with the vacuum environment may make it somewhat challenging
to observe the resulting
linear and angular velocities, but estimates of terminal angular
velocities on chiral bodies are promising.
 For dielectric-metal bodies, the force is toward the metal side,
due to the low emissivity and high reflectivity of the metal.  The 
corresponding torque is in the same sense.

Many intriguing theoretical ideas have emerged in the last few years
concerning forces and torques exerted on small bodies by exotic surfaces
and even by the pervasive blackbody radiation.  The thermal vacuum acts 
as a non-trivial medium, against which a body with suitable anisotropy can move
or rotate.  This has been a subject totally dominated by theory; however,
ideas are now emerging which we hope in the next decade can reveal 
experimental signatures of these dynamical nonequilibrium Casimir effects.

\begin{acknowledgements}
This work was supported in part by a grant from the US National Science
Foundation, number 2008417.  We thank
Xin Guo,  Prachi Parashar, and Steve Fulling
for collaborative assistance.  This paper
reflects solely the authors' personal opinions and does not represent
the opinions of the authors' employers, present and past, in any
way. For the purpose of open access, the authors have applied a
CCBY public copyright license to any Author Accepted Manuscript version
arising from this submission.\end{acknowledgements}

.

\end{document}